\documentstyle[12pt]{article}

\begin{document}
\title{\bf Cosmological and Astrophysical Bounds
on Neutrino Masses and Lifetimes\thanks{Contribution to the Franklin
    Institute Symposium honoring Fred Reines.
    An extension of an
    earlier paper appearing in Phys. Rev {\bf D45}, 4720 (1992).
    Supported in part by
    DOE Contract No. DOE-AC02-76-ERO-3071.}}
\author{{SIDNEY A. BLUDMAN}\\
{\it Department of Physics, University of Pennsylvania,
Philadelphia, PA 19104}\\}
\date{UPR-530-T}
\maketitle

\begin{abstract}

The best upper bounds on the masses of stable and unstable light
neutrinos derive from the upper bound on the total mass density, as
inferred from the lower limit $t_0> 13$ Gyr
on the dynamical age of the Universe: If the Universe is
matter-dominated,
$m_\nu<35(23)\times$ max$[1, (t_0/\tau_\nu)^{1/2}]$ eV, according as
a cosmological constant is (is not) allowed.
The best constraints on the radiative decay of light neutrinos derive
from the failure to observe prompt gamma rays accompanying the neutrinos
from Supernova 1987A: For any $m_{\nu} > 630$  eV, this
provides a stronger bound on the neutrino transition moment than that
obtained from red giants or white dwarfs.  For
$m_\nu > 250$ eV or $\tau_\nu < t_{rec}\sim 7\times 10^{12}$ sec,
the upper limit on the radiative branching ratio is even smaller than
that obtained from the limits on $\mu$-distortion of the cosmic
background radiation.
Our results improve on earlier cosmological
and radiative decay constraints by an overall factor twenty, and
allow neutrinos more
massive than 35 eV, only if they decay overwhelmingly
into singlet majorons or other new particles.
\end{abstract}

\setcounter{footnote}{1}
\section
{Mass Limits on Stable and Unstable Neutrinos from the Age of the
Universe}  The masses of stable neutrinos, $\Sigma m_{\nu_i}
= 92 \Omega_{0 \nu} h^2$ eV, are bounded by $\Omega_{0 \nu} h^2
< \Omega_0 h^2$, the total cosmological mass density in units
of $\rho _{CR} h^{-2} = 10.54~\mbox{keV cm}^{-3}$.  The best
constraint on  $\Omega_0 h^2$ does not come from poorly-known
limits on  $\Omega_0$ and the Hubble constant $H_0=100h~\mbox{km s}^{-1}
\mbox{ Mpc}^{-1}$ separately, but from the present dynamical
age of the  universe, believed to be $t_0 =  (13-17)~$Gyr.
Allowing the generous limits  $H_0 = (50-100)\mbox{ km s}^{-1}\mbox{ Mpc}
^{-1},~ 0.66  <  H_0 t_0  <  1.7$.

If the Universe is now matter-dominated and there is no
cosmological constant, then the age limits allow $\Omega_0
\leq 1,~\Omega_0 h^2 \leq 0.25,~m_\nu\leq 23~eV ^{1}$.
If the dimensionless
cosmological constant $\lambda_0 \equiv \rho_{vac}/\rho_{CR}
\leq 0.8^{2}$, the age limits allow $\Omega_0 < 1.5,~
\Omega_0 h^2 < 0.38,~m_\nu\leq~35 eV$.  If
we imposed the flat space condition $\Omega_0 +
\lambda_0 = 1$ or allowed a presently
radiation-dominated Universe, much lower bounds would be obtained.
Nevertheless, we conservatively adopt the upper bounds $m_\nu\leq 35 (23)
{}~eV$ for
any stable neutrino with (without) a cosmological constant.
Massive neutrinos can
evade this cosmological bound only if they either decay
into relativistic products  (a light neutrino and either a photon
or Goldstone boson) at red-shift
$1 + z_D > m_\nu / (92 \Omega_0h^2)$.  (Annihilation into a pair of
majorons is fast enough, only if the $B-L$
symmetry breaking takes place at a very low ($< 50$ MeV) scale.$^3$)

This redshift is achieved at time
$t= (1.7)\frac{1}{2} (\Omega_0 h_0^2)^{-1/2}(1 + z_D)^{-2} \\
< (7.1 \times 10^4~\mbox{Gyr})
(\Omega_{0 \nu} h^2)^{3/2} (m_{\nu}/\mbox{eV})
^{-2}$.$^4$.  Thus the proper lifetime of the decaying neutrino,

$$ \tau_{\nu} = 2.1 \times 10^{21} (\Omega_{0 \nu} h^2)^{3/2}
( m_{\nu}/\mbox{eV})^{-2}
< 5.0 \times 10^{20}(m_{\nu}/\mbox{eV})^{-2},\eqno(1)  $$
where hereafter all times are in
seconds, all masses in eV.  This constraint on $\tau_{\nu}$
is shown by the horizontally shaded region at the top of
Fig. 1, along with the minimum dynamical age $\tau_0 = 13$ Gyr.
This argument conservatively
assumes nothing more than that radiation is
red-shifted in an expanding universe.

This decay of a massive neutrino
would leave the Universe radiation-dominated.
In order to allow a later
matter-dominated epoch long enough for
the evolution of
large-scale structure, a massive neutrino must decay even earlier,
with $\tau_{\nu} < 1$ yr.$^{5}$  This stronger lifetime limit
depends on theories for the evolution of large-scale structure, which
we need not assume.

\section {Cosmological Bounds on Radiative Decays}
Massive neutrino decay into photons must be slow enough
that the decay photons not unacceptably distort the cosmic
background radiation spectrum accurately observed by COBE.
{}From
the observed limit on $\mu$-distortion,  following
Altherr et al$^6$, we obtain the radiative branching ratio limits

$$B_\gamma \leq \left\{    \begin{array} {lr}
1.4\times 10^{7} {\tau_\nu}^{-2/3}/m_\nu \leq 0.012/m_\nu,
  &  \tau_\nu < t_{rec}\\
9.1\times 10^{-15}\tau_\nu,       &  \tau_\nu > t_{rec}.
\end{array} \right.     \eqno(2)
$$
Here $B_\gamma$ is the branching ratio into photons, and
we have taken $T_\gamma(t_{rec})=0.308~\mbox{eV},~t_{rec}=4.39\times
10^{12}({\Omega_0}h^2)^{-1/2}$ for the epoch of radiation-matter
recombination.

Direct searches of the ultra-violet background
for photons
entering our Galaxy with red-shifted energies below the hydrogen
ionization threshold$^7$, already show that, provided absorption by
dust may be neglected,
the radiative branching ratio must be
$B_\gamma <10^{-6}-10^{-5}$.

\section {Bounds on Radiative Decay from Supernova 1987A}
The strongest constraint on radiative decays derives generally
from the failure of the Solar Maximum Mission Gamma Ray Spectrometer
(GRS)$^{8}$ to detect any prompt gammas
from Supernova 1987A in the Large Magellanic Cloud at distance
$\tau_{LMC} = 5.7 \times 10^{12}$ light seconds from Earth.$^{9}$.
Underground neutrino detectors observed an
electron antineutrino fluence on Earth
$\phi_{\bar{\nu_e}} = (8 \pm 3) \times 10^9 ~ \bar{\nu}_e \mbox{ cm}
^{-2}$,
emitted from a $\bar{\nu_e}$-neutrinosphere
at temperature T = 4.2~MeV and mean energy $<E _{\bar{\nu}_e}>$ =
12.5 MeV.  Because
$\nu_{\mu,\nu}$ and $\bar{\nu}_{\mu,\tau}$ were emitted
from deeper in the star, at  T = 8 MeV with average energy
$<E_{\nu_\tau}>$ = 25 MeV and half the fluence of $\bar{\nu}_e$,
the combined fluence of $\nu_{\tau} + \bar{\nu}_{\tau}$ must have been
$\approx \phi_{\bar{\nu_e}}$.

For a T = 8 MeV Fermi-Dirac spectrum of decaying
$\nu_{\mu,\tau}$, a fraction $W_{\gamma}$= 0.6 of the decay
photons would fall in the detector's 10-25 MeV window.  If F is the
fraction of $\nu_{\tau}$ that decay before reaching us and $B_{\gamma}$
the branching ratio into photons, then the expected gamma fluence
$\phi_{\nu} F W_{\gamma} B_{\gamma} < \phi_{{\gamma} max}$.  The SSM GRS
was maximally sensitive, $\phi_{\gamma max}=0.4$ photons $cm^{-2}$,
to those gammas that might have arrived within 10 sec after the
$\bar{\nu}_e$ pulse, because the decaying neutrino was light or decayed
fast enough.  In this way we obtain the bounds

$$B_\gamma F\leq \left\{ \begin{array} {lr}
2\times 10^{-12}m_\nu &  50 < m_{\nu_\tau} < 250~\mbox{eV} \\
8\times 10{-11}      &
   m_{\nu_\tau} > 250,~<50~\mbox{eV},  \end{array} \right.  \eqno(3)
$$
where $F$ is the fraction of $\nu$ decaying before reaching Earth.  Here
$\tau_{LAB} = \tau_{\nu} (E/m)$ is the massive neutrino's lifetime in the
laboratory,
$\tau_{\nu}$ is its proper lifetime, and
$t^{\ast} = 20(E/m_{\nu})^2$ is the time within which
a massive neutrino must decay
in order that its light decay products arrive no later than 10 sec after
the $\bar{\nu}_e$ burst. Thus,
$t_{LMC}/\tau_{LAB} = 2.4 \times
 10^5 m_{\nu}/\tau_{\nu},~t^{\ast}/\tau_{LAB}
= 5.0 \times 10^8/\tau_{\nu}m_{\nu}$ for 25 MeV $\nu_{\tau}$.

Comparison with eq. (2) then shows two things:

(1) If the neutrino lifetime is short enough ($\tau_{\nu} < 2.4 \times 10^5
m_{\nu} < 6 \times 10^4~\mbox{sec for }\\
m_{\nu} < 250$ eV or $\tau_{\nu} < 5 \times
10^8 m_{\nu}^{-1} < 2 \times 10^6$ sec
for $m_{\nu} >~250$ eV), then all
neutrinos decay before reaching Earth (F = 1) and the radiative branching
ratio must be very small, $B_{\gamma} < 10^{-10}$ for $m_{\nu} < 50$ eV
or $>$
250~eV, $B_{\gamma} < 2 \times 10^{-12} m_{\nu}$ for $50 < m_{\nu} <
250$ eV;

(2) If
$\tau_{LAB} > t_{LMC}$ or $t^{\ast}$, the radiative decay
lifetime must be long
$$ \tau_{\nu}/B_{\gamma} > \left\{
\begin{array}  {ll}
2.8 \times 10^{15} m_{\mu} &               m_{\nu} < 50\mbox{ eV} \\
1.4 \times 10^{17}&                       50 < m_{\nu} < 250~\mbox{eV} \\
6.0 \times 10^{18} m_{\nu}^{-1} &         50\mbox{eV} < m_{\nu} .
\end{array}
\right. \eqno(4)$$

The region of radiative decay lifetime
excluded (assuming $B_{\gamma} > 10^{-10}$), is shown by the
vertical shading in Fig. 1.  Our constraint is four times stronger than
that in ref. (9) because we have
corrected the neutrino average energy and the gamma fraction, energy and
fluence limit to be that of $\nu_{\tau}$ rather than $\nu_e$.

Parametrizing the radiative decay rate $B_\gamma/\tau_\nu = \mu^2 m_\nu
^3/8 \pi = 5.16 (\mu/\mu_B)^2 m_{\nu}^3$ by a transition magnetic moment
$\mu$, for $m_{\nu} > 250$ eV, Eq. (4) requires in Bohr magnetons,
$\mu/\mu_B < 1.8 \times 10^{-10} m_{\nu}^{-1}$.
{}From the absence of fast cooling of white dwarf or red giant stars by the
transverse plasmon decay into $\nu+ \bar{\nu}$, we already
know$^{10}$ that $\mu < 3 \times 10^{-12} \mu_B$, but only
for $m_{\nu} < 10$ keV. Because these stars are essentially at
temperature 10 keV, the plasmon mass and the mass of any decay products
is kinematically constrained to $<$ 10 keV.
For $m_{\nu} > 630$  eV, the SMM limits on SN1987A gamma rays therefore
provides a stronger bound on the neutrino transition moment than is
obtained from red giants or white dwarfs.

Our four-fold improvement in the
SMM bound and five-fold improvement in
the cosmological bound (1) together require that
any unstable neutrino decay with radiative branching ratio $B_\gamma< 4
\times 10^{-4}$ for $m_{\nu} < 250$ eV and $<9 \times 10^{-6}
m_{\nu}$ for $m_{\nu} > 250$ eV.  The cosmological bound (2) from the
absence of $\mu$-distortion is stronger only when both
$m_\nu < 250$ eV and $\tau_\nu < t_{rec}$, in which case $B_\gamma <
5.6 \times 10^{4}\tau_\nu^{-2/3}$.  A
massive neutrino can exist only if it
decays predominantly by non-radiative (invisible) decay modes.
\listoffigures

FIG. 1. Neutrino masses and lifetimes that are excluded cosmologically,
by the age of the Universe  $(\Omega_oh^2 \leq 0.38)$ and astrophysically,
by the absence of gamma rays accompanying the Supernova 1897A
neutrinos.  The latter constraint plotted
is on $\tau _{\nu}/B_{\gamma}$, the
radiative decay lifetime, and shows that a neutrino of mass between
35 eV and 40 MeV can exist only if it decays superfast by exotic
(nonradiative) modes.

\end{document}